\documentclass[12pt]{article}

\usepackage{epsfig}
\usepackage{cite}
\def\Order#1{${\cal O}(#1)$}

\makeatletter

\def\paragraph{\@startsection{paragraph}{4}{\z@}{+2.00ex plus
 +1ex minus +.2ex}{1.5ex plus .2ex}{\it\normalsize}}

\def\section{\@startsection {section}{1}{\z@}{+3.0ex plus +1ex minus
  +.2ex}{2.3ex plus .2ex}{\normalsize\bf\boldmath}}
\def\subsection{\@startsection{subsection}{2}{\z@}{+2.5ex plus +1ex
minus +.2ex}{1.5ex plus .2ex}{\normalsize\bf\boldmath}}
\def\subsubsection{\@startsection{subsubsection}{3}{\z@}{+3.25ex plus
 +1ex minus +.2ex}{1.5ex plus .2ex}{\normalsize\bf\boldmath}}

\expandafter\ifx\csname mathrm\endcsname\relax\def\mathrm#1{{\rm #1}}\fi

\newcounter{saveeqn}

\@addtoreset{equation}{section}

\begin{document}

\vspace*{1cm}
\begin{center}
{\Large \bf  
  Theoretical error of luminosity cross section at LEP$^{\star}$ }

\vspace*{1cm}

{\sc S. Jadach}

\vspace*{.5cm}

{\normalsize \it
  Institute of Nuclear Physics,  \\
   Krak\'ow, ul. Radzikowskiego 152, Poland}
\par
\end{center}

\vspace{20mm}
\begin{center}
\bf Abstract
\end{center} 
{
  The aim of this note is to characterize briefly
  main components of theoretical error of the small angle Bhabha
  measurement at LEP
  and to discuss critically how solid these estimates really are,
  from todays perspective.
  We conclude that the existing theoretical error of the
  LEP luminometer process (small angle Bhabha) is rather solid,
  and we add some new discussion concerning the remaining uncertainties
  and prospects of the future improvements toward the $\le 0.025\%$
  precision.
}

\vspace{20mm}
\begin{center}
  Invited talk presented at Mini-Workshop 
  {\it ``Electroweak Physics Data and the Higgs Mass''},
  DESY Zeuthen, Germany, February 28 - March 1, 2003.
\end{center}

\vfill
\vspace{30mm}
\footnoterule
\noindent
{\footnotesize
\begin{itemize}
\item[${\star}$]
Work partly supported
by the European Community's Human Potential
Programme under contract HPRN-CT-2000-00149 ``Physics at Colliders'',
by Polish Government grants
KBN 5P03B09320 
and 2P03B00122, 
\end{itemize}
}

\section{The role of the luminosity measurement in the SM fits to LEP data}

\begin{figure}[!ht]
\label{fig:NuNumb}
\begin{center}
  \epsfig{file=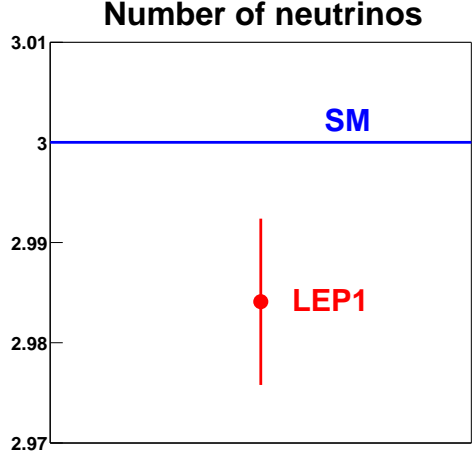,width=70mm}\\
  \vspace{-10mm}
\end{center}
\caption{\sf\small
  Numer neutrino from LEP experiments form ref...
}
\end{figure}
The luminosity measurement enters directly or indirectly into all LEP measurables.
However, as pointed out in the introduction article of this collection,
its role is not critical for the overall fit of the Standard Model to the data,
except the measurement of the invisible width on the $Z$ resonance,
which is usually formulated as the measurement of the ``neutrino'' number $N_\nu$
and is presently quoted to be $N_\nu=2.9841\pm 0.0083$,
see ADLO summary paper~\cite{Abbaneo:2001ix}
and recent review \cite{Renton:2002wy}.
This result is illustrated in Fig.~\ref{fig:NuNumb}.
$N_\nu$ deviates $1.9\sigma$ from the SM value which is normalized to exactly three,
taking into account radiative corrections to the partial widths.
The two main ingredients in the measurement of $N_\nu$ are the cross section at
the $Z$ resonance peak (here the luminosity enters) and the strong coupling
constant%
\footnote{ Alternatively one may say that there are strong correlations
  between $N_\nu$ and $\alpha_s$ in the overall fit.} $\alpha_s$.
The pure experimental error of the luminosity measurement at LEP is around 0.04\%,
the best for the OPAL experiment, a remarkable 0.034\%,
see ref.~\cite{Abbiendi:1999zx}.
The total error of luminosity at LEP is dominated by the theoretical error,
due to uncertainty in the theoretical prediction for the
luminometer QED process, Bhabha scattering in the small angles $3^\circ$-$6^\circ$,
which is the main subject of this article.

\subsection{Determination of $N_\nu$}
Let us sketch briefly how the invisible width and $N_\nu$ parameters are
determined in the LEP experiments.
Let us define the invisible decay rate of $Z$ as follows
\begin{equation}
  R_{\rm inv} \equiv \frac{ Br(Z\to {\rm inv}) }{ Br(Z\to l^+l^-)}.
\end{equation}
In the SM the invisible branching ratio $Br(Z\to {\rm inv})$ of $Z$ is identified with
its decay into three neutrinos.
The pole cross section for $e^-e^+\to Z\to f\bar{f}$
can be written in terms of the partial widths
\begin{equation}
 \sigma_{ff}^{pole}
 = \frac{12\pi}{m_Z^2} \frac{\Gamma_{e^+e^-}\Gamma_{f\bar f}}{\Gamma_Z^2}
 = \frac{12\pi}{m_Z^2} Br(Z\to e^+e^- )Br(Z\to f\bar f).
\end{equation}
The invisible branching ratio is, of course, equal the total
branching ratio (equal one) minus the branching ration for decay
into hadrons and all charged leptons
\begin{equation}
 Br(Z\to {\rm inv}) = 1- Br(Z\to {\rm had}) - 3 Br(Z\to l^+l^-).
\end{equation}
The invisible decay rate is conveniently reformulated in terms of the
visible leptonic pole cross sections $\sigma_{l^+l^-}^{pole}$ and the
ratio of the hadronic to leptonic pole cross sections $R_{\rm had}$
\begin{equation}
 R_{\rm inv} \equiv \frac{ Br(Z\to {\rm inv}) }{ Br(Z\to l^+l^-)}
  = \bigg(  \frac{12\pi}{m_Z^2 \sigma_{l^+l^-}^{pole}}\bigg)^{1/2}  -R_{\rm had} -3.
\end{equation}
The Standard Model value of the invisible branching ratio,
taking into account radiative corrections, is
\begin{equation}
 R_{\rm inv}^{\rm SM} \equiv \frac{ 3 Br(Z\to \nu \bar \nu ) }{ Br(Z\to l^+l^-)}
  = 5.973 \pm 0.003, \quad \hbox{ for} \quad \alpha_s = 0.119\pm 0.003,
\end{equation}
which we compare with the experimental value.
Traditionally the ratio of the two is referred to as
the experimentally measured ``neutrino number''
\begin{equation}
 N_\nu = 3 \frac{R_{\rm inv}}{ R_{\rm inv}^{\rm SM}} = 2.9841 \pm 0.0083.
\end{equation}
Note that the QED ISR theoretical error 
in $\sigma_{l^+l^-}^{pole}$ of $\pm 0.02\%$ is too small
to contribute significantly to error of $N_\nu$.

\section{Theoretical prediction for the luminometer process}

\begin{figure}[!ht]
\begin{center}
  {\epsfig{file=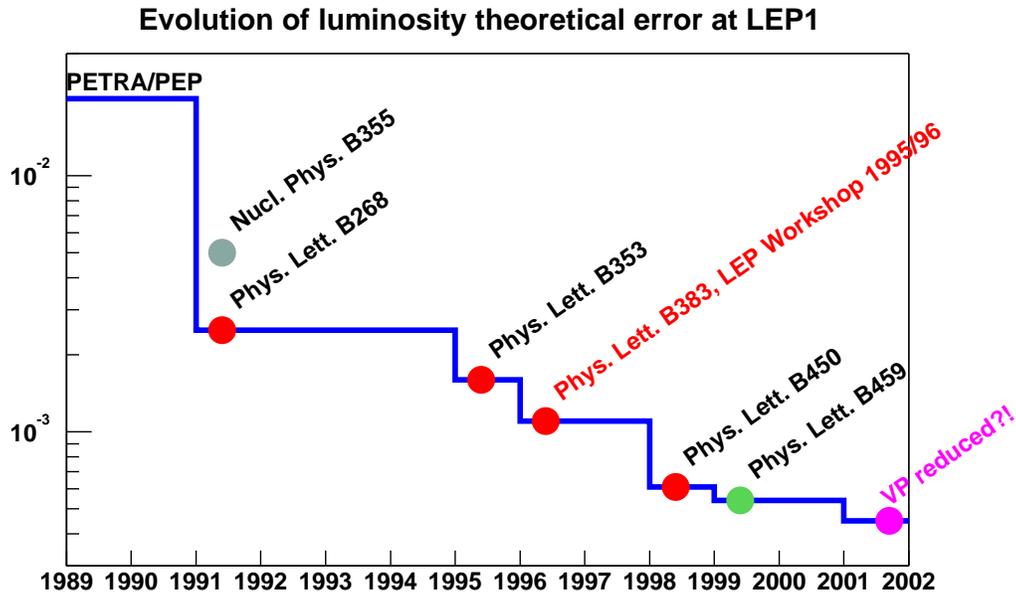,width=140mm}}\\
  \vspace{-10mm}
\end{center}
\caption{\sf\small
  Time evolution of the theoretical luminosity error at LEP era.
\label{fig:LumiEvol}
}
\end{figure}

All LEP collaborations calculate theoretical predictions for the luminometer
small angle Bhabha process using BHLUMI Monte Carlo program version 4.04
of ref.~\cite{Jadach:1997is}.
Theoretical uncertainty of the BHLUMI prediction,
mainly due to unaccounted higher order corrections,
is however provided/estimated by a series of auxiliary works, see below.

The time evolution of the theoretical error in the LEP luminosity
measurement is depicted in an approximate way in Fig.~\ref{fig:LumiEvol}.
This plot reflects error estimates from papers
\cite{Beenakker:1991es,Jadach:1991zf,Jadach:1991pj,Jadach:1995pd,Arbuzov:1996eq,%
      Ward:1998ht,Montagna:1999eu}.
At the PEP/PETRA times the best prediction for the small angle Bhabha process was
provided by the MC program of Kleiss and Berends
of ref.~\cite{Berends:1983fs} and its error was estimated
to be $\sim 2\%$.
The beginning of LEP was addressed by
the improved calculation of ref.\cite{Beenakker:1991es} with
the precision tag of $0.5\%$.
In the parallel works of refs.~\cite{Jadach:1991zf,Jadach:1991pj}
the new precision $0.25\%$ was achieved
and this precision was referring for the first time to the prediction of 
the multiphoton BHLUMI Monte Carlo 
of refs.~\cite{Jadach:1989ec,Jadach:1992by}.
The later improvements of the BHLUMI has led to its version of ref.~\cite{Jadach:1997is}
and its new theoretical precision $0.15\%$ was established first
in ref~\cite{Jadach:1995pd}.
The LEP workshop 95/96 \cite{Jadach:1996gu} was ``The great consolidation'',
which has led to a new precision level%
\footnote{This precision is meant for the prediction of BHLUMI 4.04~\cite{Jadach:1997is}.}
of $0.11\%$.
It was based on the comparison of several calculations, revising all components of the theory
error. The final result was published in 
the joint paper~\cite{Arbuzov:1996eq} of the workshop participants.
After the 95/96 workshop there was a major reduction of the uncertainty
due to photonic corrections in ref.\cite{Ward:1998ht},
giving total precision $0.061\%$,
and another one of ref.~\cite{Montagna:1999eu},
where new  estimation of the light pairs contribution
has lead to an even lower error of 0.054\%.
The last entry Fig.~\ref{fig:LumiEvol} corresponds to
``speculation'' of the present note in which
the recent reduction of the error of the hadronic vacuum polarization
contribution $\delta$VP=0.40$\to$0.025\% provides for total theoretical error of 0.045\%.

\begin{figure}[!ht]
\begin{center}
  {\epsfig{file=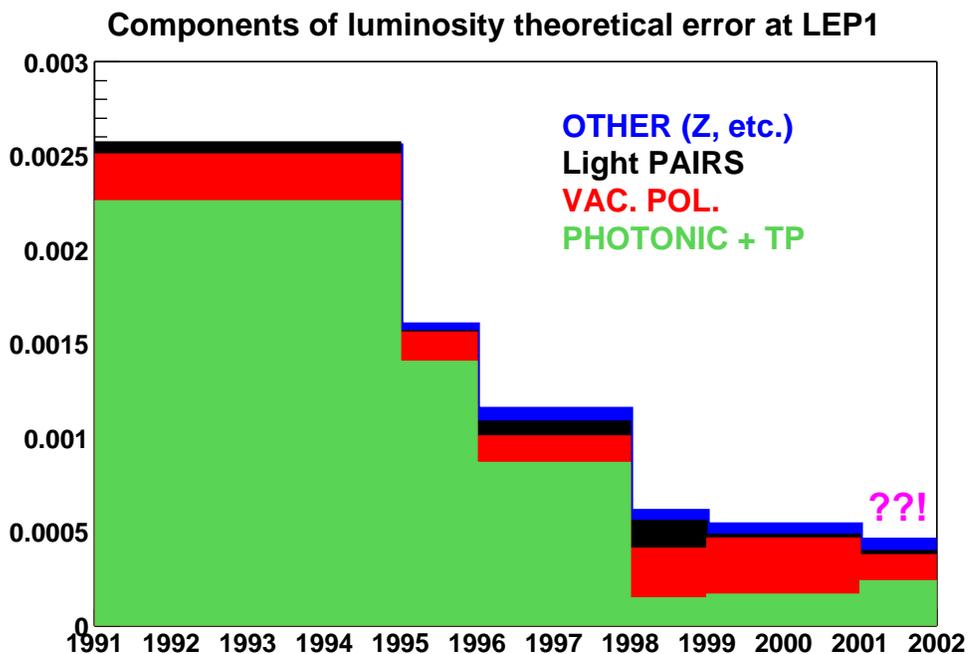,width=140mm}}\\
  \vspace{-10mm}
\end{center}
\caption{\sf\small
  Components of the theoretical luminosity error and 
  their time evolution.
}
\label{fig:LumiComp}
\end{figure}

\subsection{Main components of the theoretical luminosity error}
The main components of the theoretical luminosity error and 
their time evolution are depicted%
\footnote{The plotted quantity is the total error which is split
  into components proportionally to the square of the component.
  This justified by the fact that components are combined in the quadrature.}
in Fig.~\ref{fig:LumiComp}.
As we see, the dominant component until work of ref.~\cite{Ward:1998ht}
was the missing photonic correction, mainly second order subleading.
For the reasons discussed below, the {\em technical precision}%
\footnote{Technical precision summarizes all kind of numerical problems,
  including bugs in the programs and mistakes in numerical algorithms.}
was in most cases combined together with the missing photonic corrections%
\footnote{Simply because there was no other MC of a comparable quality
  for establishing technical precision of BHLUMI in a solid and independent way.}.
In the last two papers~\cite{Ward:1998ht,Montagna:1999eu}, see Fig.~\ref{fig:LumiComp},
vacuum polarization contribution has become dominant,
see below for discussion about its possible reduction,
which is already indicated in Fig.~\ref{fig:LumiComp}.

\subsection{Photonic corrections}

\begin{table}[!ht]
\begin{center}
{\small
\newcommand{\mystrut}{{\hbox{\rule[-3mm]{0mm}{8mm}}}}
\def\half{ {1\over 2} }
\def\alf1{ {\alpha\over\pi} }
\newcommand{\perm}{{\times 10^{-3}}}
\begin{tabular}{||l|r|c|c|c|c||}
\hline\hline
 \multicolumn{6}{|c|}{ Canonical coefficients in PHOTONIC corrections,
     $L=\ln(-t_{\min}/m_e^2)$}
\\  \hline
    \multicolumn{2}{|c|}{ }
  & \multicolumn{2}{|c|}{  $\theta_{min}=30$~mrad  }
  & \multicolumn{2}{|c||}{ $\theta_{min}=60$~mrad   }
\\  \hline
    \multicolumn{2}{|c|}{ }
  & LEP1 &  LEP2 
  & LEP1 &  LEP2 
\\  \hline\hline \mystrut 
${\cal O}(\alpha L    )$     &    $  \alf1 4L$         
       &\bf  $137\perm$   &  $152\perm$   &  $150\perm$   &  $165\perm$  
\\  \hline\hline \mystrut 
${\cal O}(\alpha      )$     &    $2 \half \alf1 $    
       &\bf  $2.3\perm$  &  $2.3\perm$    &  $2.3\perm$   &  $2.3\perm$  
\\ \hline \mystrut 
${\cal O}(\alpha^2L^2 )$     &    $  \half \left(\alf1 4L\right)^2 $ 
       &\bf  $9.4\perm$   & $11\perm$   & $11\perm$       & $14\perm$   
\\ \hline\hline \mystrut 
${\cal O}(\alpha^2L )$       &    $ \alf1 \left(\alf1 4L\right) $ 
       &\bf  $0.31\perm$  & $0.35\perm$  & $0.35\perm$    & $0.38\perm$  
\\ \hline \mystrut 
${\cal O}(\alpha^3L^3)$      &    $ {1\over 3!}\left(\alf1 4L\right)^3 $ 
       &\bf  $0.42\perm$  & $0.58\perm$  & $0.57\perm$    & $0.74\perm$  
\\ \hline\hline
\end{tabular}}\\
\vspace{-5mm}
\end{center}
\caption{\sf\small
  Canonical coefficients determining size of the photonic corrections.
}
\label{tab:Coeff}
\end{table}

\begin{figure}[!ht]
\begin{center}
  \fbox{\epsfig{file=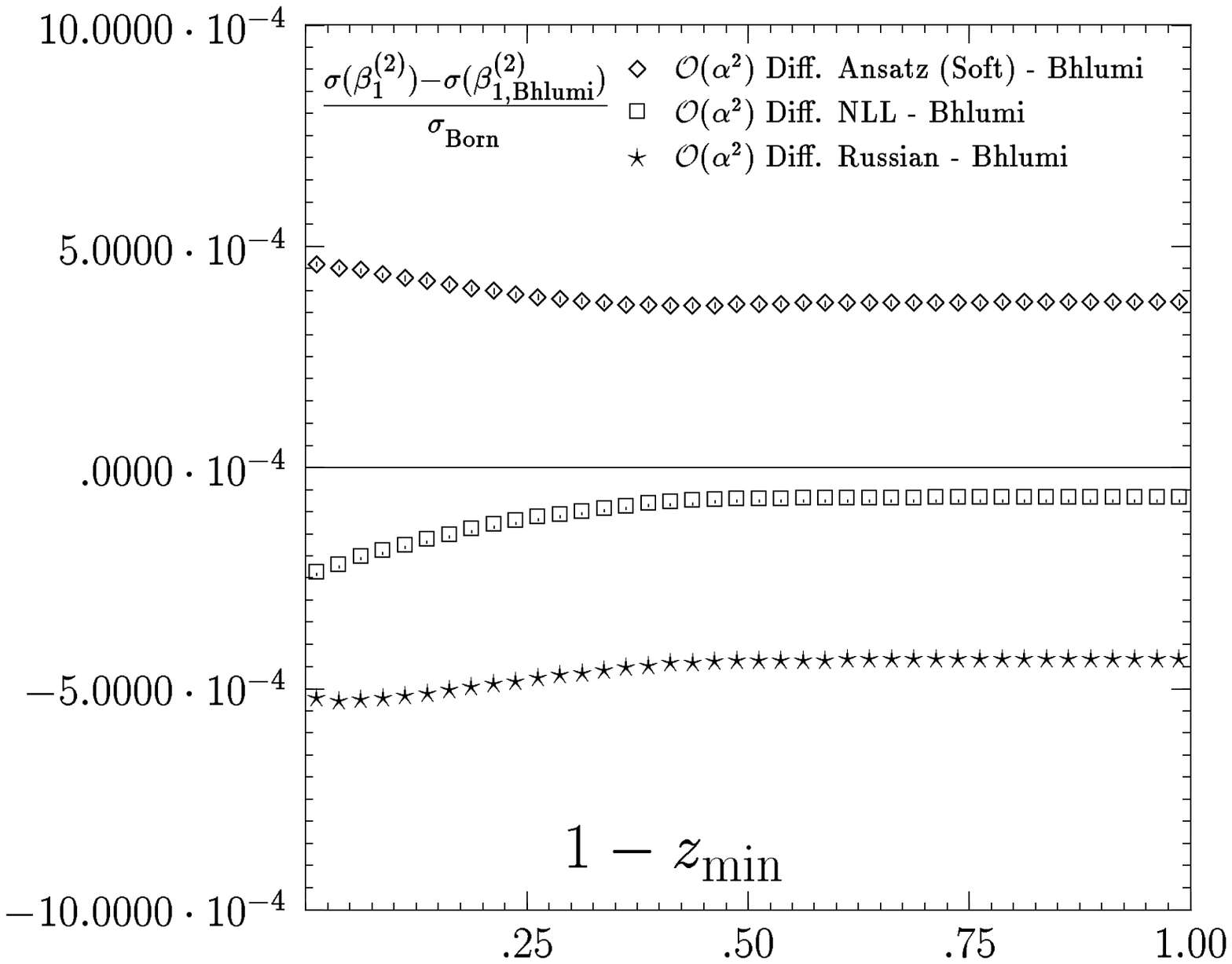,height=62mm}}
  \fbox{\epsfig{file=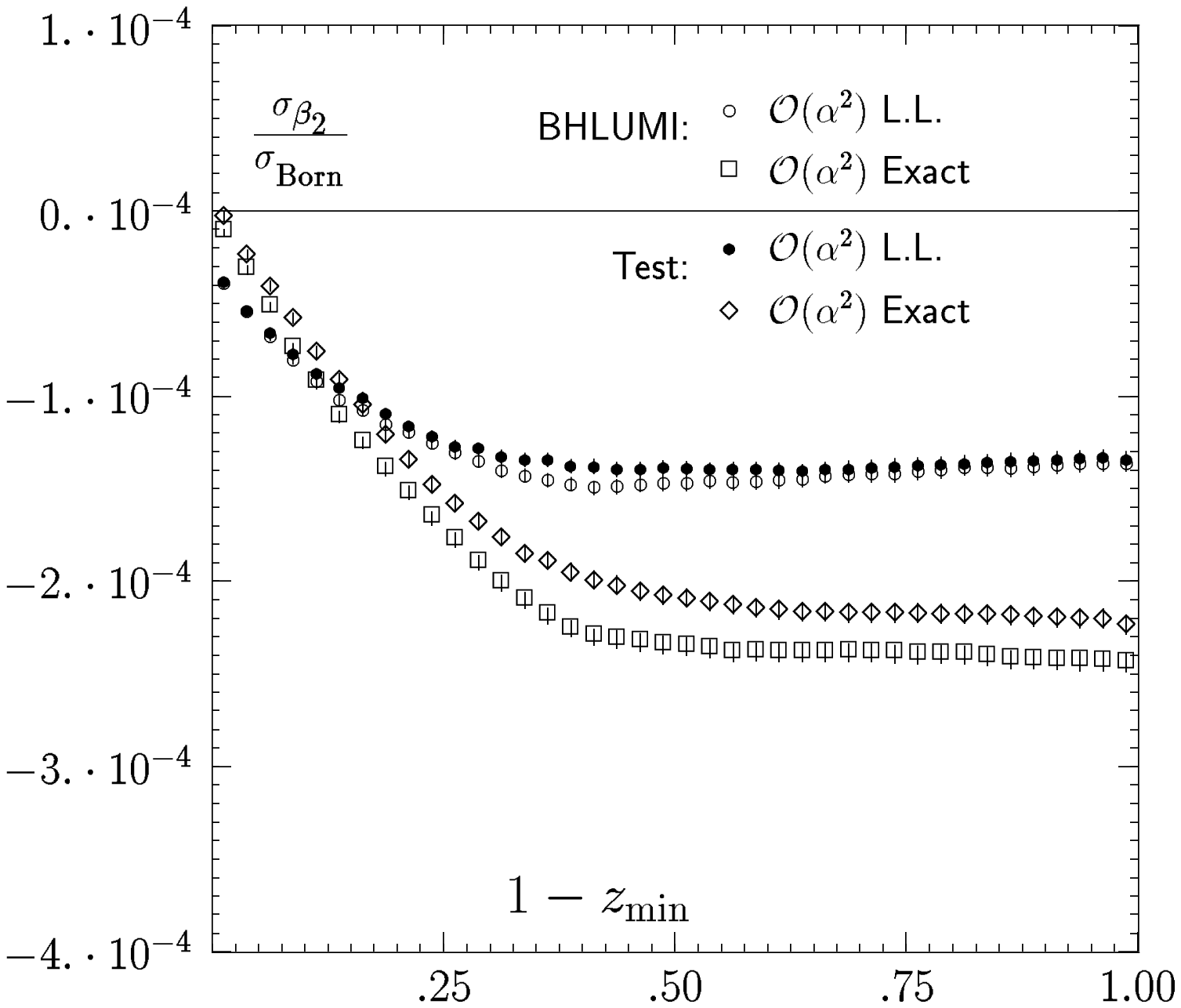,height=62mm}}\\
  \vspace{-5mm}
\end{center}
\caption{\sf\small
  Missing second order subleading contributions in BHLUMI 4.04, according to
  calculation ref.~\cite{Ward:1998ht}, for realistic LEP event selections.
}
\label{fig:Photonic}
\end{figure}
As we see from the values of the generic coefficient of the photonic
corrections presented in first column of Table~\ref{tab:Coeff},
we always expected the missing second order subleading and third order leading
photonic corrections to be below $0.1\%$.
The actual proof that it is true was provided 
in refs.~\cite{Jadach:1996ir,Ward:1998ht}.
In particular results of ref.~\cite{Ward:1998ht} were the real breakthrough.
In Fig.~\ref{fig:Photonic} we present the principal results of this paper,
which demonstrates that the second order subleading photonic correction in BHLUMI
4.04 is below%
\footnote{In fact the estimate of the total missing \Order{\alpha^2 L} in this reference
  is an overestimate, because its three components are added in the quadrature
  and again combined with technical precision, see below for more discussion.}
$0.03\%$.
Unfortunately this correction is not included in the published BHLUMI
code -- 
as a consequence, the conclusion of ref.~\cite{Ward:1998ht} is restricted
to experimental selections considered in this paper.

\subsection{Fermion pairs}
Main contribution is coming from the $e^+e^-$ pairs.
For a typical calorimetric LEP detector there are strong virtual-real cancellations
and this is why this contribution is small for the inclusive event selections
in all LEP luminometers.
Contrary to some early claims that that this contribution is huge, $\sim 0.500\%$,
in ref~\cite{Jadach:1997ca} the light pair corrections were found
to be $ -0.013\% \pm 0.020\%$, for the realistic LEP event selections.
Using similar techniques, this contribution
was calculated independently in refs.~\cite{Montagna:1998vb,Montagna:1999eu}
and for typical LEP acceptance it was found there to be
from $-.025\%$ to $-.030\%$, with the technical+physical precision $\simeq 0.010\%$.
Both groups provided MC tools for correcting experimental data%
\footnote{There exists unpublished version of BHLUMI which includes light pair production;
the corresponding version 2.30 of BHLUMI is described in ref.~\cite{Jadach:1997ca}.}.

\begin{figure}[!ht]
\begin{center}
  {\epsfig{file=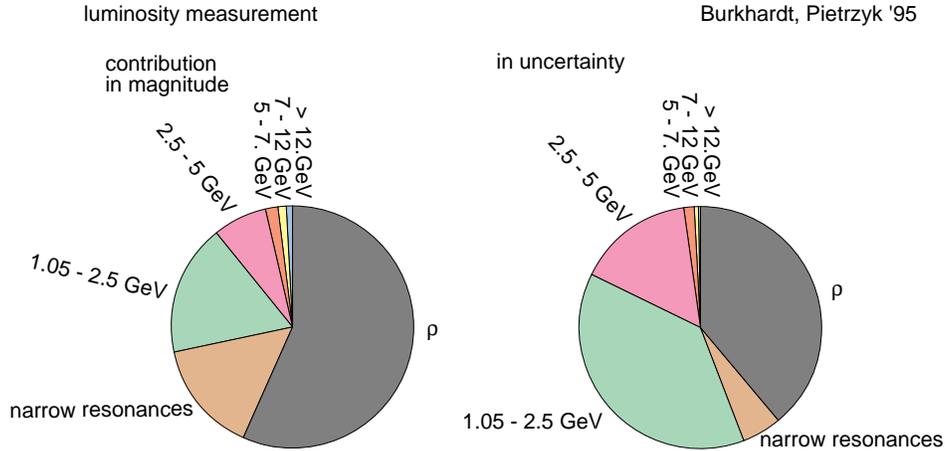,height=60mm}}\\
  \vspace{-5mm}
\end{center}
\caption{\sf\small
  vacuum polarization, from ref~\cite{Jadach:1996gu}.
}
\label{fig:VP}
\end{figure}

\subsection{Technical precision}
\label{sec:TP}
In ref.~\cite{Jadach:1997bx} the technical precision (TP)
of the unexponentiated \Order{\alpha} (no exponentiation)
was determined in the MC calculation to be $0.013\%\pm 0.017\%$.
It was calculated using  difference between MC and semi-analytical
codes, with $\pm 0.017\%$ MC statistical error.
Since work of ref.~\cite{Jadach:1997bx} we had good reasons to believe that
TP$\simeq 0.020\% -0.030\%$ for the bulk of the \Order{\alpha} 
MC small angle Bhabha cross section.

New tests of LEP 95/96 workshop, see Fig.~15 in ref.~\cite{Jadach:1996gu},
for realistic event selection provided TP$\simeq 0.030\%$ using difference of two
independent unexponentiated \Order{\alpha} MC programs.
For the multiphoton BHLUMI, its TP was incorporated into the 
estimate of the photonic theoretical  error 0.10\%,
which was estimated in ref.~\cite{Jadach:1996gu} by examining variety of the MC programs.
The size of TP alone of the multiphoton BHLUMI
was {\em guessed} to be $\simeq 0.20-0.040\%$, see Table~21 in this reference.

Let us stress that until the LEP 95/96 workshop the technical precision of BHLUMI
was always tied up together with the missing photonic correction 
(mostly \Order{\alpha^2 L}).
An attempt of determining TP of the multiphoton BHLUMI was done
in ref.~\cite{Jadach:1997bx} for the {\em quasi-realistic}%
\footnote{Event selections in this exercise were defined in terms of the 
  final electrons not combined with FSR photons and in terms of 
  the transfer $t$ instead of the final electron angles.}
event selection by comparing the BHLUMI MC with the set of special 
high quality semi-analytical calculation.
The agreement of $\simeq \pm 0.017\%$ was found, not only for the total
cross section but also separately for each of many components of the 
multiphoton QED multiphoton matrix element.

Summarizing the TP$\simeq \pm 0.030\%$ seems to be a reasonable guesestimate
of the technical precision for the multiphoton BHLUMI 4.x
for a wide range of event selection,
but we still miss some powerful test, which would definitely determine
TP of the multiphoton BHLUMI 4.x at the level of $0.01\%$,
{\em independently} of the missing higher orders.

\subsection{Hadronic vacuum polarization}
\label{sec:VP}
Until 1996-98 works of refs.~\cite{Burkhardt:1995tt,Eidelman:1995ny}
were used as a source of hadronic vacuum corrections to small angle Bhabha.
The error due to VP was  $0.040\%$,
its {\em relative} importance increased over the years.
As seen from ``pie-plot'' of Fig.~\ref{fig:VP}, it comes mainly from $R(s)$ 
in the $\rho$ region.
There, recent measurements of $R(s)$ are now much better and will be even better
in the near future.
Let us estimate how much can we profit from this  progress.
At the typical LEP luminometer average angle $\langle \vartheta \rangle = 0.034$,
ie. $\langle \sqrt{|t|} \rangle = 1.54$GeV,
using refs.~\cite{Burkhardt:1995tt} and \cite{Karlen:2001hw} we estimate now
$\Re \Pi_{1995}= 0.541\pm 0.022$ and 
$\Re \Pi_{2001}= 0.535\pm 0.014$%
\footnote{ H. Burkhardt, private communication.}.
Rescaling naively, we get the reduction of the error due to VP: $0.040\% \to 0.025\%$.
Of course, a more systematic study should be done.
Note, however, that since VP correction is rather weakly dependent on the details
of the experimental event selection, it can be redone even the after LEP data analysis
is finished%
\footnote{This might be not necessarily true for the photonic corrections.}.

\begin{table}[!ht]
\begin{center}
\begin{tabular}{||l|l|l|l||}
\hline\hline
 & \multicolumn{2}{|c|}{LEP1} & LEP2 \\
\hline
  Type of correction/error    
&   Ref.~\cite{Jadach:1995pd} &  Refs.~\cite{Jadach:1996gu,Arbuzov:1996eq}    
                              &  Refs.~\cite{Jadach:1996gu,Arbuzov:1996eq}    \\
\hline
(a) Missing photonic ${\cal O}(\alpha^2 L)$ &
    0.15\%      &\bf   0.10\%    & 0.20\% \\
(a) Missing photonic ${\cal O}(\alpha^3 L^3)$ &
    0.008\%     &\bf   0.015\%   & 0.03\%\\
(c) Vacuum polarization &
    0.05\%      &\bf   0.04\%    & 0.10\% \\
(d) Light pairs &
    0.04\%      &\bf   0.03\%    & 0.05\%\\
(e) Z-exchange  &
    0.03\%      &\bf   0.015\%   &  0.0\%\\
\hline
    Total  &
    0.16\%      &\bf   0.11\%    & 0.25\%\\
\hline\hline
\end{tabular}
\end{center}
\caption{\sf\small
  Error budget at LEP Workshop 95/96~\cite{Jadach:1996gu,Arbuzov:1996eq}, ``The Great Consolidation''
}
\label{tab:budget1}
\end{table}

\section{Total error budget}

Let us now look into and discuss in a bit more details the budget of the 
total theoretical luminosity error, in the context of the last papers in this subject.
In Table~\ref{tab:budget1} we recapitulate and compare results of
refs.~\cite{Jadach:1995pd,Jadach:1996gu,Arbuzov:1996eq}.
The work of LEP 95/96 workshop~\cite{Jadach:1996gu,Arbuzov:1996eq}
improves on almost all components, mainly on the photonic \Order{\alpha^2 L}
and revises upward the \Order{\alpha^3 L^3} component, according
to ref.~\cite{Jadach:1996ir}.
Technical precision is a part of the \Order{\alpha^2 L} quoted as $0.11\%$.
How big part? It is not specified, see discussion above.
In ref.~\cite{Jadach:1995pd} there is a suggestion that technical precision alone 
of BHLUMI is $0.02\%-0.04\%$.

\begin{table}[!ht]
\begin{center}
\begin{tabular}{||l|l|l|l|l||}
\hline
  Type of correction/error    
&  Refs.~\cite{Jadach:1996gu,Arbuzov:1996eq} 
&  Refs.~\cite{Ward:1998ht} &  Refs.~\cite{Montagna:1999eu}  & update\\
\hline
Technical precision &
     $-$   &   $-$~~~~~~~~~\footnotesize (0.03\%) 
           &   $-$~~~~~~~~~\footnotesize (0.03\%) &\bf 0.03\% \\
Missing photonic ${\cal O}(\alpha^2 L)$&
   0.10\%  &  0.027\%~~\footnotesize (0.013\%)   
           &  0.027\%~~\footnotesize (0.013\%)    &\bf 0.013\%\\
Missing photonic ${\cal O}(\alpha^3 L^3)$ &
  0.015\%  &  0.015\%~~\footnotesize (0.006\%)   
           &  0.015\%~~\footnotesize (0.006\%)    &\bf 0.006\%\\
Vacuum polarization &
  0.04\%   &  0.04\%    &  0.040\%                &\bf 0.025\%\\
Light pairs  &
  0.03\%   &  0.03\%    &  0.010\%                &\bf 0.010\%\\
Z-exchange   &
 0.015\%   &  0.015\%   &  0.015\%                &\bf 0.015\%\\
\hline
  Total    &
  0.11\%   & 0.061\%~~{\footnotesize (0.062\%)} 
           & 0.054\%~~{\footnotesize (0.055\%)}     &\bf 0.045\%\\
\hline\hline
\end{tabular}
\end{center}
\caption{\sf\small
  My personal update of LEP1 theoretical error, Febr. 2003
}
\label{tab:budget2}
\end{table}

In the second and third column of Table~\ref{tab:budget2} we present 
the next two improvements of refs.~\cite{Ward:1998ht,Montagna:1999eu}.
In the second column~\cite{Ward:1998ht} we find the
dramatic improvement of photonic part,
while in the third column~\cite{Montagna:1999eu} we see an improved value
of the light pairs.
The other entries are unchanged with respect to 
Refs.~\cite{Jadach:1996gu,Arbuzov:1996eq},
and again both paper~\cite{Ward:1998ht,Montagna:1999eu} refer to corrections
which are omitted or incomplete in BHLUMI 4.04 program.

Let us elaborate more on the estimate of the technical
precision of BHLUMI adopted in refs.~\cite{Ward:1998ht,Montagna:1999eu}.
In ref.~\cite{Ward:1998ht} a great deal of effort was done 
to get under control the TP of 
the \Order{\alpha^2 L} photonic contribution itself,
which was calculated there and which {\em is not included} in BHLUMI 4.04.
However, this does not directly address the question of the
TP of the corrections which {\em are included} in BHLUMI 4.04.
The estimate of the TP of BHLUMI 4.04 in refs.~\cite{Ward:1998ht,Montagna:1999eu}
is essentially the same as in 
refs.~\cite{Jadach:1995pd,Jadach:1996gu,Arbuzov:1996eq},
see also discussion in Sec.~\ref{sec:TP},
except that we now know that most of $0.10\%$, which was attributed to
both technical precision and \Order{\alpha^2 L} missing photonic,
is in fact not the \Order{\alpha^2 L} missing photonic!
So how big TP is?
The entire $0.10\%$ is definitely too conservative as an etsimate of the
technical precision of BHLUMI, because there is a number of partial tests
which show that \Order{\alpha^1} part is under control to within $0.02\%$
and that the other parts beyond \Order{\alpha^1} agree with a series of 
special semianalytical tests in ref.~\cite{Jadach:1997bx} to within $0.017\%$,
albeit for special event selection, see Sec.~\ref{sec:TP}.
In the following we shall adopt $0.03\%$ as a conservative
estimate of the TP of BHLUMI 4.04, separately from any QED missing corrections.

In the 2-nd and 3-rd column of Table~\ref{tab:budget2}
we show in brackets the actual missing \Order{\alpha^3 L^3} and \Order{\alpha^2 L} 
contributions according to refs.~\cite{Jadach:1996ir,Ward:1998ht},
isolating the TP of $0.03\%$ in a separate entry.
The quoted value of the missing \Order{\alpha^2 L} of $0.013\%$
corresponds to a coherent (not in quadrature) sum of contribution 
from $\beta_0$ and $\beta_1$, see Fig.~\ref{fig:Photonic}, and from $\beta_2$
(with the dominant contribution from the vertex correction in $\beta_0$) --
all that from ref.~\cite{Ward:1998ht}.
The \Order{\alpha^3 L^3} missing contribution of $0.006\%$ is taken from 
Fig. 2 in ref.\cite{Jadach:1996ir} for the SICAL detector,
without any ``safety factor'' of two.
The new total results $0.062\%$ and $0.055\%$ 
in the brackets in 2-nd and 3-rd column of Table~\ref{tab:budget2}
are almost the same as the original ones.

In the 4-th column of Table~\ref{tab:budget2}
we show also the result for the total theoretical error,
assuming again the TP of $0.03\%$, and
decreasing the vacuum polarization error to $0.025\%$, as suggested
by the exercise in Sect.\ref{sec:VP}.
The total error is now $0.045\%$, 
see bottom row of the 4-th column in the Table~\ref{tab:budget2}.
Obviously, the above exercise should be done in a more systematic way.
Nevertheless, it illustrates very well {\em the remaining uncertainties} in the game
of establishing theoretical error of the luminometer cross section for LEP1.

Last but not least let us try to answer whether it is feasible
to reduce the total theoretical
error down to $\simeq 0.025\%$ and what is necessary to achieve this goal.
The presently available CPU power allows one to compare
two MC programs with statistical error below  $0.01\%$.
The toughest problem will be to reduce the technical error for the two multiphoton
MCs, that is BHLUMI and another one, let us call it BHLUMI+, to the level of 
$0.01\%$. 
The hypothetical BHLUMI+ should feature a new QED matrix element, of
the CEEX type of ref.~\cite{Jadach:2000ir} including
the \Order{\alpha^2 L} of ref.~\cite{Ward:1998ht}
and augmented with the complete \Order{\alpha^3 L^3}.
It should necessarily feature an alternative parametrization of the multiphoton phase
space~\cite{jadach:unpubl}.
The treatment of $Z$-exchange would automatically improve due to the use of CEEX.
Hadronic VP should be updated.
Light pair corrections are already under sufficient control.
I believe that one could profit from this improvement even when LEP
data analysis is finished, i.e. it will be possible to propagate it to $N_\nu$
and $\sigma^{pole}$, because the difference between BHLUMI+ and BHLUMI
would be very small and weakly dependent on event selection.
We would then see whether the $1.9\sigma$ discrepancy with the SM in $N_\nu$
disappears or becomes even larger%
\footnote{
  NB. there are models with the massive neutrino 
  mixing~\cite{Jarlskog:1990kt,Bilenky:1990tm},
  which suggest the presence of a deficit in the experimental $N_\nu$.}.
The rest of the fits of the SM to the data would be unaffected.

\section{Summary}

\begin{itemize}
\item 
  The present theoretical error of the 
  small angle Bhabha $\simeq$0.06\% seems rather solid.
\item 
  The room for an easy improvement exists (vacuum polarization).
\item
  Radical improvement of the TH precision to the level of $\leq 0.025\%$,
  i.e. below the best experimental error $0.034\%$ is feasible.
\item
  This will require reduction of the technical precision and absorbing
  the existing improvements of the photonic QED corrections in the MC.
  Hadronic VP will get reduced another factor 2 in the meantime.
\end{itemize}

\noindent
{\bf Acknowledgments}

I am very grateful for very helpful discussion to  B.F.L.~Ward and M.~Skrzypek.


\providecommand{\href}[2]{#2}

\end{document}